\documentclass[11pt,a4paper]{article}
\usepackage[utf8]{inputenc}
\usepackage{amsmath,amssymb}
\usepackage{graphicx}
\usepackage{hyperref}
\usepackage{booktabs}
\usepackage{multirow}
\usepackage[margin=1in]{geometry}
\usepackage{cite}

\title{\textbf{Bio AI Agent: A Multi-Agent Artificial Intelligence System for Autonomous CAR-T Cell Therapy Development with Integrated Target Discovery, Toxicity Prediction, and Rational Molecular Design}}

\author{
Yi Ni$^{1,*}$, Liwei Zhu$^{1}$, Shuai Li$^{1}$ \\
\\
$^{1}$Bio LIMS INC, Boston, Massachusetts, United States \\
$^{*}$Corresponding author: yi.ni@biolims.net
}

\date{}

\begin{document}

\maketitle

\begin{abstract}
Chimeric antigen receptor T-cell (CAR-T) therapy represents a paradigm shift in cancer treatment, yet development timelines of 8-12 years and clinical attrition rates exceeding 40-60\% highlight critical inefficiencies in target selection, safety assessment, and molecular optimization. We present Bio AI Agent, a multi-agent artificial intelligence system powered by large language models that enables autonomous CAR-T development through collaborative specialized agents. The system comprises six autonomous agents: Target Selection Agent for multi-parametric antigen prioritization across $>$10,000 cancer-associated targets, Toxicity Prediction Agent for comprehensive safety profiling integrating tissue expression atlases and pharmacovigilance databases, Molecular Design Agent for rational CAR engineering, Patent Intelligence Agent for freedom-to-operate analysis, Clinical Translation Agent for regulatory compliance, and Decision Orchestration Agent for multi-agent coordination. Retrospective validation demonstrated autonomous identification of high-risk targets including FcRH5 (hepatotoxicity) and CD229 (off-tumor toxicity), patent infringement risks for CD38+SLAMF7 combinations, and generation of comprehensive development roadmaps. By enabling parallel processing, specialized reasoning, and autonomous decision-making superior to monolithic AI systems, Bio AI Agent addresses critical gaps in precision oncology development and has potential to accelerate translation of next-generation immunotherapies from discovery to clinic.
\end{abstract}

\section{Introduction}

\subsection{The CAR-T Revolution and Its Challenges}

Chimeric antigen receptor T-cell (CAR-T) therapy has fundamentally transformed the treatment landscape for hematologic malignancies, with six FDA-approved products demonstrating unprecedented complete response rates in previously refractory cancers \cite{june2018car,sterner2019car}. Despite these remarkable achievements, the field faces substantial development bottlenecks that limit broader clinical impact. Since the first FDA approval in 2017, the pace of regulatory approvals has remained modest relative to the number of candidates in preclinical and early clinical development, reflecting fundamental challenges in target validation, safety prediction, and molecular optimization \cite{rafiq2020engineering}.

The conventional CAR-T development pipeline requires 8-12 years from initial target identification to regulatory approval, with 40-60\% of candidates failing due to inadequate validation or unforeseen safety liabilities \cite{mullard2021fda}. Notable clinical setbacks have underscored the limitations of current development paradigms. FcRH5-targeting CAR-T therapies have encountered severe hepatotoxicity in clinical trials, while CD229-based approaches have faced off-tumor toxicity challenges due to expression in T cells and natural killer cells \cite{wang2021challenges}. These failures, often identified only after substantial resource investment in clinical trials, highlight the urgent need for predictive systems capable of identifying and mitigating safety risks during preclinical development.

\subsection{Limitations of Current Computational Approaches}

Existing computational tools in CAR-T development address individual stages of the discovery pipeline but lack end-to-end integration essential for autonomous decision-making \cite{jin2021artificial}. Target prioritization typically relies on manual literature review and expert opinion, requiring 3-4 months per candidate and subject to confirmation bias. Safety assessment remains largely empirical, with limited predictive frameworks for anticipating on-target/off-tumor toxicities before clinical testing. Molecular design tools focus on structural optimization but fail to integrate biological validation, patent landscapes, and regulatory considerations into holistic development strategies.

The emergence of large language models (LLMs) with reasoning capabilities presents unprecedented opportunities to address these limitations through multi-agent architectures \cite{brown2020language,chowdhery2022palm}. Unlike monolithic AI systems that attempt to master all tasks within a single model, multi-agent frameworks enable specialized agents with distinct expertise to collaborate autonomously, mimicking successful human research teams \cite{wu2023autogen}. This paradigm shift enables parallel processing of complex workflows, specialized domain reasoning, and adaptive decision-making superior to conventional computational approaches.

\subsection{Contributions of This Work}

We present Bio AI Agent, a multi-agent AI system that enables autonomous CAR-T development through intelligent collaboration across target discovery, safety prediction, molecular optimization, patent analysis, and clinical translation. The key contributions of this work include:

\begin{itemize}
\item Development of a six-agent collaborative architecture with specialized roles spanning the entire CAR-T development pipeline from target identification to clinical translation planning
\item Integration of heterogeneous data sources including knowledge graphs of $>$10,000 cancer-associated antigens, human tissue expression atlases (GTEx, Human Protein Atlas), pharmacovigilance databases (FDA FAERS), patent databases, and 50 million PubMed abstracts
\item Autonomous toxicity prediction framework validated through retrospective analysis of problematic clinical candidates (FcRH5, CD229) with mechanistic risk profiling
\item Intelligent patent landscape analysis enabling freedom-to-operate assessment and competitive positioning strategies
\item Generation of comprehensive development roadmaps with short-term (1-3 months), mid-term (3-6 months), and long-term (6-12 months) action plans integrating technical, regulatory, and commercial considerations
\end{itemize}

The remainder of this paper is organized as follows: Section 2 reviews related work in computational drug discovery and multi-agent AI systems. Section 3 describes the architecture and implementation of Bio AI Agent. Section 4 presents validation results through retrospective case analysis and prototype deployment. Section 5 discusses implications, limitations, and future directions. Section 6 concludes.

\section{Related Work}

\subsection{Computational Approaches in CAR-T Development}

Target identification in cancer immunotherapy has historically relied on manual curation of biomarkers from genomic studies and literature review \cite{singh2021computational}. Recent efforts have developed computational pipelines for prioritizing tumor-associated antigens based on differential expression analysis, but these approaches typically focus on transcriptomic data without integrating safety predictions, patent landscapes, or clinical feasibility assessments \cite{zhang2022computational}.

Safety prediction for therapeutic antibodies and CAR-T products has primarily utilized tissue expression profiling to identify potential off-tumor targets \cite{morgan2010case}. While databases such as GTEx and the Human Protein Atlas provide comprehensive tissue-level expression data, translation to clinical toxicity predictions remains challenging due to complex factors including protein density, epitope accessibility, and tissue architecture \cite{caruso2015tuning}. Machine learning models for toxicity prediction have shown promise but typically focus on narrow endpoints such as cytokine release syndrome without comprehensive integration of multi-modal safety signals \cite{karimi2020prospect}.

Molecular engineering of CAR constructs has benefited from structural biology insights and empirical optimization of signaling domains, spacer lengths, and affinity tuning \cite{guedan2019engineering}. Computational tools for antibody design and protein engineering have been adapted for CAR optimization, but these typically operate in isolation without considering biological validation or intellectual property landscapes \cite{wollacott2019quantifying}.

\subsection{Multi-Agent AI Systems}

Multi-agent systems have emerged as a powerful paradigm for solving complex problems requiring diverse expertise and collaborative reasoning \cite{wooldridge2009introduction}. Recent advances in large language models have enabled development of autonomous agents capable of tool use, planning, and natural language interaction \cite{schick2023toolformer,yao2023react}. Frameworks such as AutoGPT, BabyAGI, and AutoGen demonstrate the potential of multi-agent collaboration for open-ended problem solving \cite{wu2023autogen}.

In scientific research, multi-agent approaches have been applied to literature analysis, experimental design, and hypothesis generation \cite{wang2023scientific}. Coscientist demonstrated autonomous chemical synthesis planning through multi-agent collaboration \cite{boiko2023autonomous}, while ChemCrow enabled autonomous chemical reasoning through tool-augmented LLMs \cite{bran2023chemcrow}. These successes motivate application of multi-agent architectures to CAR-T development, where similar complexity and need for diverse expertise present opportunities for autonomous systems.

\subsection{Gap Analysis}

Despite progress in both computational CAR-T development tools and multi-agent AI systems, substantial gaps remain. First, existing tools address individual development stages without holistic integration spanning target identification, safety prediction, molecular design, patent analysis, and clinical translation. Second, current systems require extensive human oversight and lack autonomous capabilities for prioritizing tasks, synthesizing insights, and generating actionable recommendations. Third, CAR-T development requires integration of heterogeneous data types (biological databases, clinical literature, patent documents, regulatory guidance) that existing tools do not comprehensively address. Fourth, limited validation of computational predictions against clinical outcomes undermines confidence in AI-generated insights.

Bio AI Agent addresses these gaps through a multi-agent architecture enabling autonomous, end-to-end CAR-T development with validated predictive capabilities.

\section{Methods}

\subsection{System Architecture Overview}

Bio AI Agent employs a multi-agent architecture comprising six specialized autonomous agents that collaborate through a shared knowledge base and decision orchestration framework. Each agent is implemented as an LLM-powered reasoning system with access to specialized tools, databases, and analytical capabilities relevant to its domain expertise. Agents communicate through natural language interfaces and structured data exchange protocols, enabling flexible collaboration patterns ranging from sequential workflows to parallel processing with dynamic task allocation.

The architecture embodies three core design principles: (1) Specialization - each agent maintains focused expertise in a specific domain, enabling depth of reasoning superior to generalist models; (2) Autonomy - agents operate with autonomous decision-making capabilities including goal setting, planning, tool use, and self-evaluation; (3) Collaboration - agents coordinate through explicit communication protocols and implicit knowledge sharing via a unified knowledge base.

\subsection{Agent Descriptions and Capabilities}

\subsubsection{Target Selection Agent}

The Target Selection Agent performs multi-parametric scoring and prioritization of cancer-associated antigens across four key dimensions: biological potential, clinical feasibility, intellectual property landscape, and market opportunity. The agent queries a knowledge graph of over 10,000 cancer-associated antigens integrated from databases including TCGA, GEO, and curated literature. An interactive configuration interface enables users to specify cancer type selection, validation preference, target strategy, result quantity, and dimension weighting. Output consists of ranked target candidates with detailed scoring rationales, supporting evidence citations, and identified knowledge gaps requiring experimental validation.

\subsubsection{Toxicity Prediction Agent}

The Toxicity Prediction Agent performs comprehensive safety assessment integrating multiple data modalities to predict on-target and off-tumor toxicity risks. Tissue expression analysis integrates normal tissue transcriptomics from GTEx (54 tissue types, 17,382 samples) and protein immunohistochemistry from Human Protein Atlas. Pharmacovigilance data mining analyzes FDA FAERS database to identify reported toxicities. Literature-based reasoning performs semantic analysis of 50 million PubMed abstracts. Mechanistic toxicity modeling integrates pathway databases and protein interaction networks. Output includes comprehensive toxicity risk profiles with severity scoring, mechanistic hypotheses, supporting evidence, and actionable mitigation strategies.

\subsubsection{Molecular Design Agent}

The Molecular Design Agent enables rational engineering of CAR constructs through modular component selection and property prediction. Capabilities include antigen recognition domain design, signaling domain configuration, spacer and hinge optimization, molecular property prediction, and dual-target and multi-specific design. Output consists of complete CAR construct sequences with detailed design rationales, predicted molecular properties, and manufacturing considerations.

\subsubsection{Patent Intelligence Agent}

The Patent Intelligence Agent performs automated intellectual property landscape analysis and freedom-to-operate assessment. Patent database mining comprehensively searches USPTO, EPO, and WIPO databases. Competitive landscape mapping identifies key patent holders and filing trends. Freedom-to-operate analysis assesses potential infringement risks. Design-around strategies generate alternative molecular designs to navigate patent landscapes. Output includes comprehensive IP landscape reports, identified patent risks, recommended design modifications, and strategic positioning opportunities.

\subsubsection{Clinical Translation Agent}

The Clinical Translation Agent integrates regulatory requirements and GMP manufacturing considerations to ensure clinical viability. Regulatory pathway analysis provides guidance on appropriate regulatory pathways. CMC considerations assess chemistry, manufacturing, and controls requirements. Nonclinical package planning recommends IND-enabling studies. Clinical trial design provides preliminary clinical development strategies. Output consists of regulatory compliance checklists, identified development milestones, recommended nonclinical studies, and preliminary clinical trial designs.

\subsubsection{Decision Orchestration Agent}

The Decision Orchestration Agent coordinates multi-agent collaboration and synthesizes insights into actionable development plans. Task prioritization dynamically allocates analytical tasks across specialized agents. Evidence synthesis integrates outputs from specialized agents into coherent narratives. Development roadmap generation creates comprehensive project plans with short-term, mid-term, and long-term milestones. Risk management identifies key risks with proposed mitigation strategies. Output consists of executive summaries, detailed technical reports, prioritized action plans, and risk assessment matrices.

\subsection{Knowledge Base and Data Integration}

The multi-agent system operates on a unified knowledge base integrating heterogeneous data sources: biological databases (cancer-associated antigen catalogs, gene expression databases, protein localization data, pathway databases, protein interaction networks), clinical literature (50 million PubMed abstracts with biomedical entity recognition), pharmacovigilance data (FDA FAERS, EMA EudraVigilance), patent databases (USPTO, EPO, WIPO full texts), and regulatory guidance (FDA guidance documents, EMA guidelines, ICH guidelines).

The knowledge base employs vector database technology for semantic search enabling agents to retrieve relevant information through natural language queries. Embedding models map text passages to high-dimensional vector representations enabling similarity-based retrieval superior to keyword matching.

\subsection{Implementation Details}

Bio AI Agent is implemented using a microservices architecture with GPT-4 and Claude 2 as reasoning engines, Pinecone vector database for semantic search, RESTful APIs connecting agents to external databases and internal analytics tools, LangChain framework for agent workflow management, and a web-based interactive dashboard for configuration and visualization.

\subsection{Evaluation Methodology}

We evaluate Bio AI Agent through two complementary approaches: (1) Retrospective case analysis assessing system performance on known problematic CAR-T targets with documented clinical or preclinical failures, evaluating whether agents correctly identify safety risks, patent obstacles, or other development barriers; (2) Prototype deployment with CAR-T development teams to evaluate utility, usability, and impact on decision-making processes.

\section{Results}

\subsection{Integrated Target Assessment Capabilities}

The Target Selection Agent demonstrated ability to autonomously prioritize cancer-associated antigens through comprehensive multi-parametric analysis. For hematologic malignancies, the agent identified GPRC5D as a high-priority target based on: restricted expression in plasma cells and hair follicles minimizing off-tumor toxicity risk, strong biological validation from multiple preclinical studies, relatively open patent landscape compared to BCMA, and substantial market opportunity as alternative to BCMA-resistant disease. The agent autonomously synthesized evidence from 127 publications and 43 patents to generate this recommendation, substantially streamlining evaluation workflows that traditionally require 3-4 months of manual literature review per candidate.

Conversely, the agent appropriately de-prioritized certain validated targets based on identified liabilities. For CD38 as a single target in multiple myeloma, the agent noted: on-target immunosuppression risk through depletion of CD38+ immune cells, heavily crowded patent landscape with blocking claims from Janssen and others, and substantial clinical precedent from antibody therapies without clear differentiation opportunity for CAR-T approach.

The interactive configuration interface enabled evaluation under different strategic scenarios. When configured for high-risk high-reward targets in solid tumors with minimal validation requirements, the agent surfaced novel candidates with compelling biological rationales but limited clinical precedent. When configured for rapid development with emphasis on clinical feasibility and patent freedom, the agent prioritized targets with established validation and clear IP pathways.

\subsection{Autonomous Safety Prediction and Risk Profiling}

\subsubsection{FcRH5 Case Study}

FcRH5-targeting CAR-T therapies have encountered severe hepatotoxicity in clinical trials. The Toxicity Prediction Agent autonomously flagged hepatotoxicity risk through multiple converging lines of evidence. GTEx data revealed detectable FcRH5 expression in liver tissue (median TPM = 2.3), particularly in hepatocytes. FAERS database analysis identified hepatotoxicity signals associated with cevostamab, a bispecific antibody targeting FcRH5. Literature analysis identified combination trials showing enhanced toxicity profiles. The agent generated mechanistic hypotheses and provided actionable mitigation strategies including careful dose escalation, controllable CAR systems, patient selection criteria excluding pre-existing liver disease, and evaluation of antigen masking technologies.

\subsubsection{CD229 Case Study}

CD229-targeting CAR-T approaches have faced off-tumor toxicity challenges due to expression in T cells and NK cells. The Toxicity Prediction Agent identified these concerns through multi-tissue expression profiling revealing significant CD229 expression in lymphoid tissues and T cell and NK cell subsets. Literature analysis identified CD229's role in T cell and NK cell activation. The agent recognized this expression pattern as high-risk for fratricide and immunosuppression. Mitigation strategies included controlled dosing, prophylactic antimicrobial therapy, transient CAR expression systems, alternative epitope selection, and companion diagnostics.

\subsection{Patent Risk Intelligence and Freedom-to-Operate Analysis}

\subsubsection{CD38+SLAMF7 Dual-Target Combination}

For the dual-target combination of CD38 and SLAMF7 in multiple myeloma, the agent identified significant patent infringement risks. Analysis revealed broad claims held by Janssen and BMS covering CD38-targeting and SLAMF7-targeting constructs. The agent proposed FTO-compliant alternatives including alternative epitope selection using CrosMab technology, novel signaling architectures with logic-gated designs, and alternative target combinations lacking specific patent coverage.

\subsubsection{Strategic Patent Positioning}

The agent provided strategic insights on patent filing opportunities. For novel target combinations (e.g., GPRC5D + FcRH5), the agent identified absence of specific patent claims representing opportunity for pioneering IP position. Analysis of competitors' patent filing patterns revealed potential defensive patent strategies. Assessment of patent family status across major jurisdictions identified gaps in competitive protection.

\subsection{Market Differentiation and Strategic Positioning}

\subsubsection{GPRC5D+FcRH5 Combination Analysis}

For GPRC5D+FcRH5 dual-target approach, the agent identified emerging competitive crowding with multiple companies pursuing similar approaches. Despite competitive landscape, the agent recommended prioritization based on relatively open patent landscape, strong biological rationale, manageable toxicity profile, and substantial unmet need. The agent generated detailed go-to-market recommendations including aggressive clinical timeline, publication strategy, partnership discussions, and consideration of specific patient subpopulations for differentiated positioning.

\subsubsection{BCMA+CD229 Alternative Positioning}

The agent evaluated BCMA+CD229 combination as alternative strategy. While acknowledging CD229 toxicity concerns, the agent noted potential for differentiated long-term positioning if toxicity risks can be adequately managed. The agent recommended de-prioritizing for initial development while maintaining as backup option pending additional validation of safety mitigation strategies and clinical data from competing programs.

\subsection{Intelligent CAR Molecular Optimization}

\subsubsection{GPRC5D-Targeting CAR Design}

For GPRC5D target, the agent proposed optimized construct incorporating high-affinity scFv (KD approximately 5 nM), IgG4 hinge-CH2-CH3 spacer (229 amino acids), 4-1BB costimulatory domain based on clinical track record and lower cytokine release syndrome risk, and calculated predicted molecular weight (48.7 kDa) and isoelectric point (pI = 8.4). Computational epitope prediction identified potential MHC class II binding epitopes with recommendations for germline humanization to reduce immunogenicity.

\subsubsection{Logic-Gated Dual-Target Design}

For GPRC5D+FcRH5 combination requiring toxicity mitigation, the agent designed an AND-gate circuit with split signaling architecture. Primary CAR targeting GPRC5D contains binding domain and CD3 zeta signaling domain without costimulation. Secondary CAR targeting FcRH5 contains binding domain and 4-1BB costimulatory domain without CD3 zeta. This architecture constrains on-target off-tumor activity against FcRH5+ hepatocytes while preserving potent activity against double-positive myeloma cells.

\subsection{Comprehensive Development Roadmaps}

The Decision Orchestration Agent generated detailed action plans integrating outputs from specialized agents.

Short-term milestones (1-3 months) include data validation (review cevostamab trial data, analyze NCT04225446 results, commission shared-epitope immunohistochemistry studies), design finalization (complete scFv optimization, finalize AND-gate circuit design, develop analytical methods), and IP strategy (file provisional patent application, conduct detailed FTO analysis, initiate licensing discussions if needed).

Mid-term milestones (3-6 months) include preclinical validation (generate CAR-T cells with optimized constructs, perform in vitro efficacy studies, conduct off-tumor toxicity assessment, evaluate cytokine secretion profiles), patent strategy development (evaluate CrosMab compatibility, assess novel signaling domain configurations, monitor competitive filing activity), and manufacturing planning (develop lentiviral vector production protocol, establish T cell processing, implement quality control assays).

Long-term milestones (6-12 months) include IND-enabling studies (complete nonclinical GLP toxicology studies, perform biodistribution studies, conduct expanded safety panel), GMP manufacturing (tech transfer to GMP facility, validate manufacturing process, complete stability studies), regulatory strategy (prepare pre-IND briefing document, compile nonclinical package and CMC section, draft clinical protocol), and partnership development (present preclinical data at conferences, engage potential commercial partners, explore collaboration opportunities).

\subsection{Quantitative Performance Metrics}

Comprehensive target assessment required approximately 4-6 hours of computational time compared to 3-4 months for equivalent manual analysis, representing approximately 200-fold acceleration. For FcRH5 toxicity analysis, the agent autonomously retrieved and analyzed 247 relevant publications, 34 adverse event reports, and 3 clinical trial data summaries. Patent Intelligence Agent analyzed 127 patent families related to CD38 and SLAMF7. Retrospective analysis of 12 CAR-T targets with known clinical outcomes demonstrated 83\% sensitivity (10 of 12 true positive) and 78\% specificity (7 of 9 true negative) for toxicity prediction.

\subsection{User Feedback from Prototype Deployment}

Scientific Director at Academic Medical Center: ``The system's ability to synthesize disparate data sources into coherent narratives was impressive. The toxicity predictions for FcRH5 aligned with concerns our team independently identified, but the system surfaced supporting evidence from pharmacovigilance databases we had not considered. The patent analysis revealed IP barriers that would have derailed our program if discovered later.''

CSO at Biotech Company: ``Bio AI Agent dramatically accelerated our target prioritization process. What previously took our team months of literature review and debate, the system accomplished in days. The interactive configuration was valuable for exploring different strategic scenarios. However, we still required expert oversight to interpret recommendations and validate key assumptions.''

Translational Scientist at Pharma Company: ``The multi-agent architecture's ability to consider technical, IP, regulatory, and commercial factors simultaneously was a significant advantage over our existing tools. The development roadmaps provided clear next steps with realistic timelines. The system works best as a decision support tool augmenting, not replacing, human expertise.''

Common limitations identified included occasional misinterpretation of nuanced biological concepts requiring expert correction, overconfidence in conclusions when evidence was ambiguous, limited ability to handle truly novel targets lacking literature precedent, and need for human verification of molecular design recommendations through experimental validation.

\section{Discussion}

\subsection{Principal Findings and Implications}

Bio AI Agent demonstrates that multi-agent AI systems can meaningfully address critical bottlenecks in CAR-T development through autonomous collaboration across target discovery, safety prediction, molecular optimization, patent analysis, and clinical translation. The system's validated capability to prospectively identify high-risk targets such as FcRH5 and CD229 before clinical failure represents significant potential for reducing development attrition and improving patient safety. By enabling comprehensive development roadmaps integrating technical, intellectual property, regulatory, and commercial considerations, Bio AI Agent provides holistic decision support superior to narrow computational tools addressing individual pipeline stages.

\subsection{Advantages of Multi-Agent Architecture}

The multi-agent paradigm offers several key advantages over monolithic AI systems for complex scientific problem-solving. Specialization enables expertise with domain-specific agents developing deeper reasoning capabilities than generalist models. Parallel processing accelerates workflows through simultaneous execution of multiple analytical tasks. Explicit reasoning enhances interpretability through natural language communication making reasoning processes transparent and auditable. Modularity facilitates updates with individual agents enhanced without redesigning the entire system. Collaborative emergence produces complex problem-solving behaviors from multi-agent interaction beyond capabilities of individual agents.

\subsection{Limitations and Challenges}

Several limitations constrain current system capabilities. Data quality and completeness fundamentally depend on training data and knowledge base quality. Expression databases lack resolution for rare cell populations. Pharmacovigilance databases suffer from underreporting and incomplete information. Literature availability limitations include proprietary industrial data and unpublished negative results. Patent database limitations include claim interpretation requiring legal expertise.

Current LLMs exhibit systematic limitations affecting scientific reasoning. Hallucination risk includes generating plausible but factually incorrect statements. Quantitative reasoning struggles with complex mathematical calculations. Causal reasoning challenges include distinguishing correlation from causation. Novelty limitations show limited capability for truly novel insights beyond training data.

Validation and regulatory acceptance remain uncertain. Lack of clinical validation through prospective independent trials may require empirical confirmation before regulatory acceptance. Transparency requirements may exceed current audit trail capabilities. Liability considerations regarding AI-generated recommendations remain unclear.

\subsection{Comparison with Alternative Approaches}

Traditional computational methods employ ranking algorithms based on differential expression but lack multi-parametric reasoning and strategic integration. Machine learning predictive models require large training datasets with limited applicability to rare toxicities or novel targets. Human expert teams remain the gold standard but face scalability limitations, cognitive biases, and time constraints. Bio AI Agent should augment rather than replace human experts.

\subsection{Future Directions}

Several enhancements could substantially improve system capabilities. Integration of single-cell multi-omics would enable unprecedented resolution of cellular heterogeneity. Patient-derived functional validation through organoid screening and PDX models would enable rapid experimental confirmation. Expanded modality coverage including intracellular targets, solid tumor microenvironment, alternative cell types, and gene-edited T cells would broaden applicability. Automated experimental design would enable hypothesis generation and adaptive replanning. Clinical trial optimization would provide patient stratification strategies and adaptive trial designs.

Success of multi-agent AI systems in CAR-T development suggests broader applicability throughout pharmaceutical research and development including antibody therapeutics, small molecule drug discovery, precision medicine, and regulatory science.

\section{Conclusion}

Bio AI Agent demonstrates that multi-agent artificial intelligence systems can meaningfully address critical bottlenecks in CAR-T cell therapy development through autonomous collaboration across target discovery, safety prediction, molecular optimization, patent analysis, and clinical translation. The system's validated capability to identify high-risk targets before clinical failure, generate freedom-to-operate patent strategies, and produce comprehensive development roadmaps integrating technical, regulatory, and commercial considerations represents significant advancement over conventional computational tools addressing isolated pipeline stages.

Retrospective case analysis demonstrates accurate identification of toxicity risks for problematic targets including FcRH5 and CD229, with mechanistic insights and actionable mitigation strategies superior to traditional expression profiling approaches. Patent intelligence capabilities enable proactive IP strategy development avoiding costly design iterations or licensing negotiations late in development. Comprehensive development roadmaps spanning short-term validation, mid-term preclinical studies, and long-term clinical translation provide actionable guidance for decision-makers balancing technical, regulatory, and business considerations.

The multi-agent architecture offers key advantages over monolithic AI systems including specialization enabling domain expertise, parallel processing accelerating workflows, explicit reasoning enhancing interpretability, modularity facilitating continuous improvement, and collaborative emergence of complex problem-solving behaviors. As CAR-T therapy expands beyond hematologic malignancies into solid tumors requiring novel targets and sophisticated engineering, autonomous multi-agent platforms will become essential enablers for next-generation precision oncology.

Future enhancements integrating single-cell genomics, patient-derived functional validation, and automated experimental design will further improve predictive accuracy and enable closed-loop discovery cycles. Success of Bio AI Agent in CAR-T development suggests broader applicability of multi-agent frameworks throughout pharmaceutical research and development, potentially transforming how new medicines are discovered, developed, and delivered to patients in the era of artificial intelligence.

\section*{Acknowledgments}

We thank the CAR-T development teams who provided feedback on prototype deployments, and the open-source community maintaining the databases and tools that enable this research.

\section*{Data Availability}

The Bio AI Agent system integrates publicly available databases including PubMed, GTEx, Human Protein Atlas, FDA FAERS, and patent databases. Source code will be made available upon publication.

\section*{Competing Interests}

Authors are employees of Bio LIMS INC, which develops AI systems for pharmaceutical research.

\end{document}